\documentclass[a4paper,twocolumn,11pt,accepted=2017-05-09]{quantumarticle}
\pdfoutput=1

\usepackage[numbers]{natbib}
\usepackage[utf8]{inputenc}
\usepackage[english]{babel}
\usepackage[T1]{fontenc}
\usepackage{amsmath}
\usepackage{hyperref}

\usepackage{tikz}
\usepackage{lipsum}

\usepackage{bm}

\usepackage{algorithm}
\usepackage{algpseudocode}

\usepackage{booktabs}
\usepackage{blindtext}

\usepackage{comment}
\usepackage{mathtools}
\usepackage{mathrsfs}
\usepackage{amsfonts}
\usepackage{amssymb}
\usepackage{dsfont}
\usepackage{amsthm}
\usepackage[english]{babel}
\usepackage{graphicx}
\usepackage{amsmath}

\usepackage{lipsum}
\usepackage{wrapfig}
\usepackage{float}
\usepackage{enumitem} 
\usepackage{color}
\usepackage{hyperref}
\usepackage[normalem]{ulem}
\usepackage{url}

\useunder{\uline}{\ul}{}

\hypersetup{colorlinks = true}

\def\bra#1{\ensuremath{\mathinner{\langle{#1}|}}}
\def\ket#1{\ensuremath{\mathinner{|{#1}\rangle}}}
\newcommand{\norm}[1]{\left\lVert #1 \right\rVert}

\newcommand{\needcite}[1]{\textcolor{red}{[Ref needed]}}
\DeclarePairedDelimiter{\ceil}{\lceil}{\rceil}
\usepackage{qcircuit}

\begin{document}

\title{Hamiltonian simulation in Zeno subspaces} 

\author{Kasra Rajabzadeh Dizaji}
\affiliation{School of Electrical, Computer, and Energy Engineering, Arizona State University, Tempe, Arizona 85287, USA}

\author{Ariq Haqq}
\affiliation{School of Electrical, Computer, and Energy Engineering, Arizona State University, Tempe, Arizona 85287, USA}

\author{Alicia B. Magann}
\affiliation{Quantum Algorithms and Applications Collaboratory, Sandia National Laboratories, Albuquerque, New Mexico 87185, USA}

\author{Christian Arenz}
\affiliation{School of Electrical, Computer, and Energy Engineering, Arizona State University, Tempe, Arizona 85287, USA}

\maketitle

\begin{abstract}
We investigate the quantum Zeno effect as a framework for designing and analyzing quantum algorithms for Hamiltonian simulation. We show that frequent projective measurements of an ancilla qubit register can be used to simulate quantum dynamics on a target qubit register with a circuit complexity similar to randomized approaches. The classical sampling overhead in the latter approaches is traded for ancilla qubit overhead in Zeno-based approaches. A second-order Zeno sequence is developed to improve scaling and implementations through unitary kicks are discussed. We show that the circuits over the combined register 
can be identified as a subroutine commonly used in post-Trotter Hamiltonian simulation methods. We build on this observation to reveal
connections between different Hamiltonian simulation algorithms.
\end{abstract}

\tocless\section{Introduction}
\vspace{-0.25cm}
The quantum Zeno effect is an intriguing phenomenon in quantum physics that has found a variety of applications in quantum information science \cite{franson2004quantum,paz2012zeno,patsch2020simulation,lewalle2023multi, PhysRevLett.120.020505, Nourmandipour:16, PhysRevA.98.032308, PhysRevA.92.012303, PhysRevA.99.052101, lewalle2023optimal,doi:10.1142/S0219749924500060,KUMARI2023169222}. Typically, frequent observations of a quantum system result in dynamics that become frozen \cite{misra1977zeno, facchi2008quantum}.  However, when only {parts} of a system are frequently observed, complex dynamics can arise  \cite{facchi2002quantum, burgarth2014exponential, zanardi2014coherent, zanardi2015geometry, arenz2016universal, albert2016geometry, arenz2020emerging}. In this setting, the quantum Zeno effect can be used to turn a simple, commutative evolution into an evolution capable of universal quantum computing \cite{burgarth2014exponential, arenz2016universal, blumenthal2022demonstration}. Here, we show that this observation opens up paths for designing Hamiltonian simulation techniques and for revealing connections between existing methods.

\begin{figure}
\centering
\includegraphics[width=0.95\columnwidth]{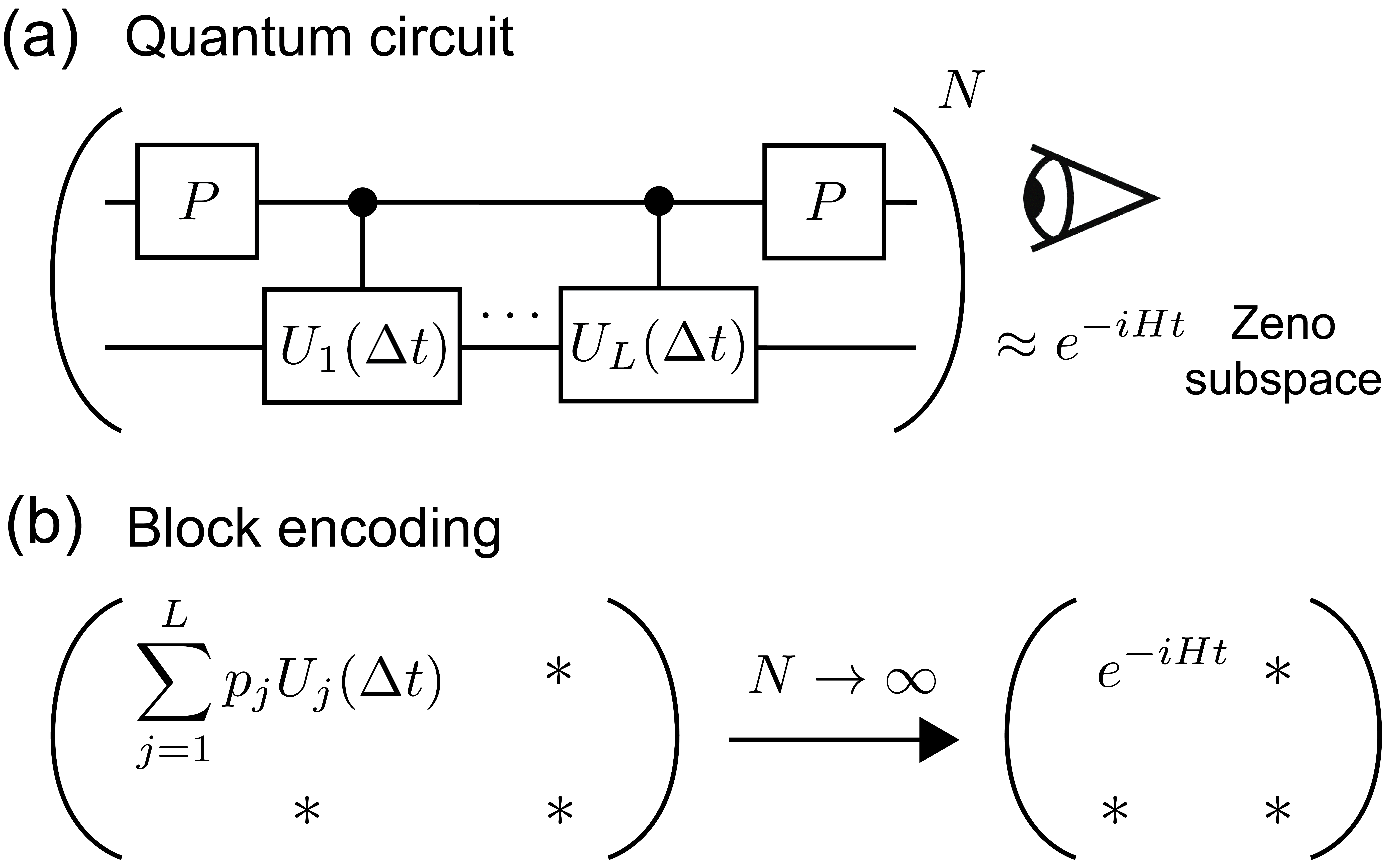}
 \caption{\label{fig:Circuits} Schematic representation of (a) the quantum circuit that is used for Zeno-based Hamiltonian simulation. The target qubit register is coupled to an ancilla qubit register through controlled operations $U_j(\Delta t)$. After a time interval $\Delta t =\frac{t}{N}$, the ancilla register is projectively measured, as described by the projector $P$. 
For sufficiently frequent measurements, the target qubit register, which constitutes a Zeno subspace, evolves according to $e^{-iHt}$ as desired. We identify one Zeno step as (b) a block encoding, $\sum_j p_j U_{j}(\Delta t)$, that approaches the unitary evolution $e^{-iHt}$ in the limit of infinitely many observations $N$.}
\end{figure}

We present Hamiltonian simulation algorithms that leverage frequent observations of ancilla qubits to simulate a desired quantum evolution, $e^{-iHt}$, on a target qubit register, as depicted in Fig. \ref{fig:Circuits} (a). A key feature of these algorithms is that in between measurements, the evolution over the combined register is commutative by construction \cite{orsucci2015hamiltonian, arenz2017lindbladian}. We show that when this commutative evolution is combined with frequent computational basis measurements of the ancilla system, the target qubit register constitutes a \emph{Zeno subspace} \cite{facchi2002quantum} in which the desired quantum evolution takes place. 
We then connect the evolution between measurements with the \textsc{select} and \textsc{prepare} subroutines commonly used in other Hamiltonian simulation algorithms, which allows us to draw on known constructions for implementing them \cite{childs2018toward, babbush2018encoding,berry2019qubitization,yuan2023optimal}. As depicted in Fig. \ref{fig:Circuits} (b), we establish connections to post-Trotter methods by identifying one Zeno step in Fig. \ref{fig:Circuits} (a) as a block encoding of a linear combination of unitaries (LCU). In the Zeno limit of infinitely fast observations of the ancilla qubits, the desired unitary evolution, $e^{-iHt}$, is obtained.

We show that the circuit complexity for this Zeno-based Hamiltonian simulation algorithm is given by $\mathcal O(D\tfrac{t^{2}\lambda^{2}}{\epsilon})$, where $D$ is the circuit complexity associated with consecutively implementing the subroutines \textsc{select} \cite{childs2018toward, babbush2018encoding, berry2019qubitization, boyd2023low, babbush2016exponentially, wan2021exponentially} and \textsc{prepare} 
\cite{babbush2016exponentially,yuan2023optimal, gui2024spacetime}, and $\lambda$ is a factor that depends on the Hamiltonian, $H$, whose corresponding time evolution over a time $t$ is being simulated. This result is similar to randomized approaches like QDRIFT  \cite{campbell2019random} that scale as $\mathcal{O}(\frac{t^{2}\lambda^{2}}{\epsilon_{\diamond}})$. The difference is that Zeno-based Hamiltonian simulation trades the classical sampling overhead in randomized approaches, which can result in additional loss of precision, for the dependence on $D$, the inclusion of ancilla qubits, and the incorporation of mid-circuit ancilla measurements (or unitary kicks). We additionally consider the prospect of higher-order Zeno sequences, in analogy to higher-order Trotter formulas \cite{childs2021theory}, to obtain improved asymptotic scaling properties, noting however that in the present work we do not attempt to match the asymptotic complexity offered by other LCU-based approaches \cite{berry2015simulating,berry2019qubitization,gilyen2019quantum}. Through the inclusion of a reflection operator, we derive a second-order Zeno formula whose circuit complexity is given by $\mathcal{O}\bigl (  D\tfrac{(\lambda t)^{1+1/2}}{\sqrt{\epsilon}} \bigr )$. Afterwards, we discuss circuit implementations of Zeno-based Hamiltonian simulation through frequent unitary kicks, rather than projective measurements. We go on to explore how the quantum Zeno effect can reveal connections and tradeoffs between different Hamiltonian simulation algorithms including QDRIFT \cite{campbell2019random}, Trotter \cite{lloyd1996universal}, and post-Trotter methods based on the LCU framework \cite{10.5555/2481569.2481570, berry2015simulating}. We conclude with a discussion of potential future directions. 

\vspace{0.5cm}
\tocless\section{Hamiltonian simulation through the quantum Zeno effect}
We consider the task of simulating time evolution under a Hamiltonian of the form 
\begin{align}
\label{eq:original}
H=\sum_{j=1}^{L} h_{j}H_{j}.	
\end{align}
We assume, without loss of generality, 
that $h_{j}>0$, and $\Vert H_{j}\Vert=1$ where $\Vert \cdot \Vert$ denotes the spectral norm. For a quantum system whose evolution is governed by $H$, the quantum Zeno effect is given by the limit \cite{facchi2008quantum,hahn2022unification}
\begin{align}
\label{eq:ZenoLimit}
\lim_{N\to \infty}\left(Pe^{-iH\Delta t}P\right)^{N}=e^{-iPHPt}P,	
\end{align}
where $\Delta t=\frac{t}{N}$ is the time interval in which projective measurements, described by the projector $P=P^{2}$, are performed. If the projector $P=\ket{\phi}\bra{\phi}$ is formed by a state $\ket{\phi}$ that is the initial state of the system, then in the Zeno limit \eqref{eq:ZenoLimit} the dynamics becomes frozen as $PHP$ becomes a phase.

To see how the quantum Zeno effect can be used for Hamiltonian simulation, we form a composite system by coupling a qubit register where the target evolution $e^{-iHt}$ will be implemented, i.e., the target register, to a register of ancilla qubits. We next construct an extended Hamiltonian $\tilde{H}$ \cite{burgarth2014exponential, orsucci2015hamiltonian, arenz2017lindbladian, arenz2016universal} that acts on this composite system and whose constituent terms mutually commute. This is achieved by coupling each $H_{j}$ in \eqref{eq:original} to a state of the ancilla register via the construction $H_{j}\rightarrow H_{j}\otimes \ket{j}\bra{j}=\tilde{H}_{j}$ \cite{arenz2017lindbladian}, such that the ancilla register contains $n_{a} = \lceil\log(L)\rceil$ qubits, and the states $\ket{j}$, $j=1,\cdots L$ form a complete and orthonormal basis for the ancilla subsystem. Throughout this work, tildes will be used to denote operators that act on the composite system, where the second factor of the tensor product corresponds to the ancilla qubit register. Omitting the coefficients $h_{j}$, the Hamiltonian on the extended space reads 
\begin{align}\label{eq:ExtendedH}
\tilde{H}=\sum_{j=1}^{L}\tilde{H}_{j}. 
\end{align}

The desired evolution $e^{-iHt}$ on the target register can then be obtained via repeated projective measurements of the ancilla qubits \cite{burgarth2014exponential,arenz2017lindbladian}, described by a projector of the form $\tilde{P}=\mathds{1}\otimes P$. We first consider $P= \ket{\phi}\bra{\phi}$, where
\begin{align}
\label{eq:preparestate}
\ket{\phi}=\sum_{j=1}^{L}\sqrt{p_{j}}\ket{j},
\end{align}
and we define $p_{j}=\frac{h_{j}}{\lambda}$ and $\lambda=\sum_{j=1}^{L}h_{j}$. Given that $\lambda \tilde{P}\tilde{H}\tilde{P}=H\otimes P$, in the Zeno limit \eqref{eq:ZenoLimit} the desired evolution is implemented over the target register as
\begin{align}
\label{eq:Zenosequence}
&\lim_{N\to \infty}\left(\tilde{P}e^{-i\lambda \tilde{H}\Delta t}\tilde{P}\right)^{N}\ket{\psi_{0}}\otimes \ket{\phi} \\
&=e^{-iHt}\ket{\psi_{0}}\otimes \ket{\phi}. \nonumber 
\end{align}
In (\ref{eq:Zenosequence}), we have assumed that the target register is initially in the state $\ket{\psi_{0}}$, while the ancilla register is initially in the state $\ket{\phi}$. Following the framework in \cite{hahn2022unification}, we show in Appendix \ref{sec:firstorder} that the corresponding Zeno error, 
\begin{align}
\epsilon=\left \Vert \left(\tilde{P}e^{-i\lambda \tilde{H}\Delta t}\tilde{P}\right)^{N}- e^{-iHt}\otimes {P}\right \Vert,	
\end{align}
is upper-bounded by, 
\begin{align}
\label{eq:ZenoError}
\epsilon\leq \frac{t^{2}\lambda^{2}}{N}.
\end{align}

Each step in the Zeno sequence \eqref{eq:ZenoLimit} is implemented with a certain probability, i.e., the probability of obtaining the measurement outcome corresponding to implementing the projector $P$. To determine this probability, we consider the state 
\begin{align}
\label{eq:stateZeno}
\ket{\varphi(\Delta t)}= 	(\mathds{1}\otimes V^{\dagger})\tilde{U}(\Delta t)(\mathds{1}\otimes V)\ket{\psi_{0}}\otimes \ket{0}^{\otimes n_{a}},
\end{align}
where $\tilde{U}(\Delta t)=\exp(-i\lambda \tilde{H}\Delta t)$ and $V$ creates the state $\ket{\phi}=V\ket{0}^{\otimes n_{a}}$.  Then, implementing the projector $\tilde{P}$ corresponds to projecting the state of the ancilla register in (\ref{eq:stateZeno}) onto $|0\rangle^{\otimes n_a}$.
We show in Appendix \ref{sec:firstorder} that the probability $p_{\text{succ}, 0}(\Delta t)=\bra{\varphi(\Delta t)}(\mathds{1}\otimes \ket{0}\bra{0}^{\otimes n_{a}})\ket{\varphi(\Delta t)}$ associated with observing $\ket{0}^{\otimes n_{a}}$ when a computational basis measurement of the ancilla register is performed is lower-bounded by $ p_{\text{succ},0} \geq 1 - 2\frac{\lambda^2t^2}{N^2}$.  We note that the sequence to prepare the state in Eq.  \eqref{eq:stateZeno} followed by the projector $\mathds{1}\otimes \ket{0}\bra{0}^{\otimes n_{a}}$ and $\mathds{1}\otimes V$ constitutes one Zeno step in Eq. \eqref{eq:Zenosequence}.

The total algorithm success probability, $p_{\text{succ}}=p_{\text{succ}, 0}^{N}(\Delta t)$, corresponding to the probability of projecting the ancilla register onto the state $\ket{0}^{\otimes n_{a}}$ over $N$ consecutive steps, thereby implementing the full Zeno sequence in Eq. \eqref{eq:Zenosequence}, can be further lower-bounded by
\begin{align}
\label{eq:success}
p_{\text{succ}}\geq 1 - 2\frac{\lambda^2t^2}{N}.	
\end{align}
As such, for a desired success probability $p_{\text{succ}}$, the number of Zeno steps $N$ required to implement $e^{-iHt}$ over the target register can be  upper-bounded by $N \leq \frac{2\lambda^{2}t^{2}}{1-p_{\text{succ}}}.$

We now consider whether there exist higher-order Zeno sequences, in analogy to higher-order Trotter formulas \cite{10.1063/1.529425,berry2007efficient, childs2021theory}, that yield a faster convergence as a function of the number of steps, $N$.   
We first observe that a Zeno sequence with a structure similar to higher-order Trotter sequences cannot succeed, as terms of the form $\tilde{P}\tilde{H}^{2}\tilde{P}$ occurring in such a sequence do not cancel. However, this problem can be circumvented by the inclusion of a unitary transformation $\tilde{R}=\mathds{1}\otimes R$, where $R=2P-\mathds{1}$ suppresses transitions to the subspace orthogonal to the Zeno subspace \cite{dhar2006preserving}. Indeed, since 
\begin{align}
\label{eq:secondorder}
&\tilde{P}\tilde{U}(\Delta t/2)\tilde{R}\tilde{U}(\Delta t/2)]\tilde{P} \\\nonumber 
&=\left(\mathds{1}-iH\Delta t-\frac{1}{2}H^{2}\Delta t^{2}\right)\otimes P+\mathcal O(\Delta t^{3}),  
\end{align}
we show in Appendix \ref{sec:secondorder} that the Zeno error for this second-order sequence is 
upper bounded by $\epsilon\leq \frac{\lambda^{3}t^{3}}{3N^{2}}$ and that a lower bound for the corresponding success probability is given by $p_{\text{succ}} \geq 1 - \frac{4\lambda^3t^3}{3N^2}$. We thus achieve a quadratic improvement by the inclusion of $R$ on the ancilla register in each Zeno step.  The prospect of developing higher-order Zeno sequences \cite{dhar2006preserving} to further improve the complexity of Zeno-based Hamiltonian simulation would constitute valuable future work and will be the subject of future investigations.

\vspace{0.5cm}
\tocless\subsection{Implementation and circuit complexity}
We provide pseudocode below for implementing $e^{-iHt}$ on the target qubit register with precision $\epsilon$ and $p_{\text{succ}}\geq 1-2\epsilon$ via $N = \lceil{t^2\lambda^2/\epsilon\rceil}$ Zeno steps. The circuit complexity is determined by the product of the number of Zeno steps, $N$, with the per-step circuit complexity $D$, where the latter corresponds to the complexity of consecutively implementing $V$, $\tilde{U}(\Delta t)$, and $V^\dagger$. Here, it is instructive to identify  $\tilde{U}(\Delta t)$ and $V$ as the \textsc{select} and \textsc{prepare} operations that are the basis of LCU-based methods, respectively, i.e., where 
\begin{align}
\label{eq:selectop}
\tilde{U}(\Delta t)=\sum_{j=1}^{L}U_{j}(\Delta t)\otimes \ket{j}\bra{j},
\end{align}
with $U_{j}(\Delta t)=\exp(-i \lambda H_{j}\Delta t)$ is the \textsc{select} operation, and where \textsc{prepare} is identified as the unitary transformation $V$ that creates the state $\ket{\phi}$ in Eq. \eqref{eq:preparestate}. The \textsc{select} and  \textsc{prepare} operations have circuit complexities of $\mathcal O(LC)$ and $\mathcal O(L)$, respectively, where $C$ is the maximum circuit complexity associated with any $U_{j}$ \cite{low2019hamiltonian}. 
Thus, in the general case, the total circuit complexity for Zeno-based Hamiltonian simulation using the first-order sequence \eqref{eq:Zenosequence} is $\mathcal O(CL\frac{t^{2}\lambda^{2}}{\epsilon})$. We note, however, that because all terms in \textsc{select} mutually commute, parallelization techniques \cite{moore2001parallel, boyd2023low,  zhang2024parallel} and techniques that leverage the structure of $H$ \cite{childs2018toward, babbush2018encoding,berry2019qubitization,yuan2023optimal, gui2024spacetime} could be explored for reducing the cost of implementation in practice. Furthermore, techniques utilizing classical optimization could also be explored to compress the controlled short-time evolution circuits \cite{PhysRevResearch.5.023146}.

\renewcommand{\algorithmicrequire}{\textbf{Input:}}
\renewcommand{\algorithmicensure}{\textbf{Output:}}

\begin{figure}
\vspace{-0.25cm}
\begin{algorithm}[H]
\renewcommand{\thealgorithm}{}
\floatname{algorithm}{}
\caption{\centering Zeno-based Hamiltonian Simulation}
\label{alg:psuedocode}
\begin{algorithmic}
\Require A Hamiltonian of the form $H=\sum_{j=1}^{L} h_{j}H_{j}$, desired evolution time $t$, and desired precision $\epsilon$.
\Ensure The implementation of $e^{-iHt}$ on a target register with precision $\epsilon$ and success probability $p_{\text{succ}}\geq 1-2\epsilon$
\State $\lambda \gets \sum_{j=1}^{L} |h_j|$
\State $N \gets \ceil{t^{2}\lambda^{2}/{\epsilon}}$ \Comment{Number of Zeno steps}
\State $\Delta t={t}/{N}$
\State Initialize ancilla register of size $n_a=\ceil{\log_2(L)}$ in $\ket{0}^{\otimes n_{a}}$
\State $i \gets 0$
\While{$i < N$}
    \State \textsc{prepare} on ancilla register
    \State \textsc{select}$(U_{j}(\Delta t))$ on the combined register
    \State \textsc{prepare}$^\dagger$ on ancilla register
    \State Computational basis measurement of ancilla register
    \State $i = i + 1$
\EndWhile
\end{algorithmic}
\end{algorithm}
\end{figure}

In implementations, a variety of resource tradeoffs can be considered, e.g., with respect to the choice of projective measurement. That is, instead of projecting onto a Zeno subspace corresponding to $P=\ket{\phi}\bra{\phi}$, a projection onto a Zeno subspace can also be achieved using $P_a = |a\rangle\langle a|$, where $|a\rangle$ is a state from a mutually unbiased basis \cite{wootters1989optimal} satisfying $\langle j|a\rangle=\frac{e^{i\phi_{j}}}{\sqrt{L}}$ for all $j=1,\cdots ,L$, where $e^{i\phi_{j}}$ is a phase. We note that if $\{\ket{j}\}$ represents the computational basis, the state $\ket{a}$ can be chosen to be an equal superposition of all computational basis states created by applying Hadamard gates to each of the ancilla qubits. If we now consider $\tilde{H}_a=\sum_{j=1}^{L}h_{j}\tilde{H}_{j}$, we have $LP_a\tilde{H}_aP_a=H\otimes P_a$. From the derivation of the Zeno error it then follows that in the setting where projective measurements are performed in a mutually unbiased basis, the Zeno error is upper bounded by $\epsilon\leq \frac{t^{2} L^{2}\Lambda^{2}}{N}$, where $\Lambda=\max_{j}|h_{j}|$. The different Zeno error scaling can be traced back to the different constructions of the projector $\tilde{P}$.

Another consideration is the use of projective measurements as a resource in the first place; in some settings, implementing repeated measurements may be undesirable or infeasible. In these settings, an alternative implementation approach could be considered where quantum Zeno dynamics are achieved by the fast application of unitary kicks \cite{PhysRevA.69.032314,burgarth2022one}, rather than projective measurements. In Appendix \ref{sec:unitarykicks} we show that for a kick operator given by $\tilde{R}$ we obtain  
 \begin{align}
\lim_{N\to \infty}(\tilde{R}\tilde{U}(\Delta t))^{N}\ket{\psi_{0}}\otimes \ket{\phi}=e^{-iHt}\ket{\psi_{0}}\otimes \ket{\phi}.
\end{align}
This corresponds to $e^{-iHt}$ being implemented over the target register with success probability $1$ in the limit of infinitely frequent kicks. For finitely many kicks, the Zeno error is of the order $\epsilon=\mathcal O(\frac{\lambda^{2}t^{2}}{N})$, which matches the complexity in Eq. (\ref{eq:ZenoError}).

\vspace{0.5cm}
\tocless\subsection{Relation to QDRIFT} 
\noindent The circuit complexity of first-order Zeno-based Hamiltonian simulation matches that of the quantum stochastic drift protocol (QDRIFT) \cite{campbell2019random} up to the circuit complexity $D$ needed to implement \textsc{select} and \textsc{prepare}. In QDRIFT, the unitaries $U_{j}(\Delta t)$ are randomly sampled according to the probability distribution $p_{j}$.  The average is determined by the quantum channel described by the completely positive and trace preserving (CPTP) map
\begin{align}
\label{eq:CPTP}
\mathcal M_{\Delta t}(\cdot)=\sum_{j=1}^{L}p_{j}U_{j}(\Delta t)(\cdot)U_{j}^{\dagger}(\Delta t).  
\end{align}
This procedure is repeated $N$ times to create the CPTP map $\mathcal M_{\Delta t}^{N}$ that describes the averaged evolution after a time $t=N\Delta t$. The more often the procedure is repeated, i.e., the larger $N$, the closer $\mathcal M_{\Delta t}^{N}$ becomes to the unitary channel $\mathcal U_{t}(\
\cdot)=e^{-iHt}(\cdot)e^{iHt}$ that describes the evolution to be simulated. The corresponding error $\epsilon_{\diamond}=\left\Vert\mathcal M_{\Delta t}^{N}-\mathcal U_{t}\right\Vert_{\diamond}$, given by the diamond distance between the two quantum channels, can be upper-bounded by $\epsilon_{\diamond}\leq 4\frac{\lambda^{2}t^{2}}{N}$ \cite{campbell2019random} (see also Appendix \ref{sec:diamond}). In the Zeno case, we achieve the repetition of the same CPTP map $\mathcal M_{\Delta t}$ by considering the state 
\begin{align}
\ket{\psi(\Delta t)}=\sum_{j=1}^{L}\sqrt{p_{j}}U_{j}(\Delta t)\ket{\psi_{0}}\otimes \ket{j},
\end{align}
before the second measurement is performed, tracing out the ancilla register, i.e., $\text{Tr}_{a}\{\ket{\psi(\Delta t)}\bra{\psi(\Delta t)}\}=\mathcal M_{\Delta t}(\ket{\psi_{0}}\bra{\psi_{0}}) $, and initializing the ancilla qubits again in $\ket{\phi}$ by measuring in the computational basis and applying the \textsc{prepare} operation. Thus, the loss of precision due to classical sampling in QDRIFT \cite{campbell2019random} is traded for the cost of executing a larger circuit over the target and ancilla register in Zeno-based Hamiltonian simulation.

\vspace{0.5cm}
\tocless\subsection{Relation to Trotter and LCU-based methods} 
Zeno-based Hamiltonian simulation is an approach derived by combining two primitives. The first primitive can be viewed, and implemented, as a first-order Trotterization of the dynamics generated by $\tilde{H}$ over a small time step $\Delta t$ to create $\tilde{U}(\Delta t) = \prod_{j=1}^Le^{-i\tilde{H}_j\Delta t}$ on the extended space. Here, the Trotterization is exact,  given that the constituent terms in $\tilde{H}$ commute. Second, projective Zeno measurements are implemented on the ancilla subsystem. This allows for approximating the target evolution, generated by a linear combination of terms $H_j$, by a linear combination of simple evolutions $U_j(\Delta t)$ on the target subspace, per Fig. \ref{fig:Circuits}(b). The result is a Hamiltonian simulation algorithm that retains the same asymptotic circuit complexity (at first order) as first-order Trotterization with respect to $t,\epsilon$, but exchanges the dependence on Hamiltonian commutators in Trotterization (see \cite{childs2021theory}) for a dependence on $\lambda$, the introduction of $\ceil{\log(L)}$ ancilla qubits, and the incorporation of mid-circuit measurements (or unitary kicks) of the ancilla register.

This combination of Trotter and Zeno primitives also allows for making connections to post-Trotter methods based on the LCU framework. In particular, as depicted in Fig. \ref{fig:Circuits} (b), one Zeno step $\tilde{P}\tilde{U}(\Delta t)\tilde{P}=\sum_{j=1}^{L}p_{j}U_{j}(\Delta t)\otimes P$ constitutes a {block encoding} \cite{low2019hamiltonian, gilyen2019quantum},
\begin{align}
&(\mathds{1}\otimes\bra{0}^{\otimes n_{a}}V^{\dagger})\tilde{U}(\Delta t)(\mathds{1}\otimes V\ket{0}^{\otimes n_{a}}) \\
&=\sum_{j=1}^{L}p_{j}U_{j}(\Delta t), \nonumber 
\end{align}
that implements $\sum_{j}p_{j}U_{j}(\Delta t)$  over the target register with probability $p_{\text{succ},0}$, for $V$ given by the \textsc{prepare} operation. The LCU utilized in Zeno-based Hamiltonian simulation is distinctive as it does not block encode the Hamiltonian, but rather, a linear combination of small-time evolutions $U_{j}(\Delta t)$ generated by the terms in the Hamiltonian \eqref{eq:original}. If these small-time evolutions are further approximated to first order in $\Delta t$ in their associated Taylor series, i.e., as $U_{j}(\Delta t)\approx \mathds{1}-i\lambda H_{j}\Delta t$ such that $\sum_{j=1}^{L}p_{j}U_{j}(\Delta t)\approx \mathds{1}-iH \Delta t$, we recover approximately the same LCU as the (first-order) truncated Taylor series algorithm \cite{berry2015simulating}, thus opening opportunities for analyzing the latter via the quantum Zeno effect. We additionally observe that in the truncated Taylor series algorithm, the success probability $p_{\text{succ}}$ is boosted through oblivious amplitude amplification \cite{10.1145/2591796.2591854} rather than frequent projective measurements of the ancilla qubits; in the present work, we utilize the same operator $R$ that is used in oblivious amplitude amplification in order to develop the second-order Zeno sequence in Eq. \eqref{eq:secondorder} that gives faster convergence.

\vspace{0.5cm}
\tocless\section{Conclusions} 
We have investigated how the quantum Zeno effect can be used for Hamiltonian simulation. To this end, we have presented a Hamiltonian simulation algorithm that is constructed from unitary evolutions, controlled on an ancilla qubit register, interspersed with frequent projective measurements of the ancilla qubits. This approach produces the desired unitary evolution on the target qubit register in the limit of infinitely fast projective measurements. Away from this limit, we have derived bounds for the associated error and success probability. We have shown that Zeno-based Hamiltonian simulation yields a circuit complexity similar to that of QDRIFT. A key difference is that the classical sampling overhead in QDRIFT is traded for executing a deeper circuit over a larger system in the Zeno-based method. We have also developed a second-order sequence that reduces the asymptotic complexity of the Zeno-based Hamiltonian simulation algorithm, yielding a quadratic improvement.  The development of $k$-th-order methods \cite{dhar2006preserving}, and more broadly, the investigation of techniques for improving the optimal asymptotic complexity, with respect to both $t$ and $\epsilon$, of Zeno-based frameworks, would constitute interesting directions for future work.

We have discussed different tradeoffs associated with the implementation of Zeno-based Hamiltonian simulation, including with respect to the choice of projective measurements and the substitution of measurements for unitary kicks. An interesting future direction could be to build on work developing hybridized algorithms for Hamiltonian simulation that aim to reduce costs by exploiting tradeoffs between different resources, e.g., the circuit (or query) complexity, the use of randomization and classical sampling, the use of classical optimization, and the inclusion of ancilla qubits \cite{childs2019faster, Rajput2022hybridizedmethods,PhysRevResearch.5.023146, hagan2023composite,chakraborty2024implementing}; frequent projective measurements could be another quantum computational resource worth consideration. We have also discussed relations to Trotter and post-Trotter methods, observing that the quantum circuits in Zeno-based Hamiltonian simulation can be implemented using the \textsc{select} and \textsc{prepare} operations, which are well-known subroutines in LCU-based approaches. Looking ahead, we posit that the quantum Zeno effect may offer a fruitful lens through which we can continue to develop and analyze quantum algorithms for Hamiltonian simulation and other applications.

We conclude by noting that our focus in this work has been on the quantum Zeno effect implemented by frequent projective measurements or unitary kicks. However, quantum Zeno dynamics can also be achieved through strong dissipative processes \cite{zanardi2014coherent, zanardi2015geometry, arenz2016universal, albert2016geometry, arenz2020emerging} or strong classical fields \cite{burgarth2022one}. Thus, other implementations of the Zeno-based Hamiltonian simulation algorithms may be worth exploring, depending on the quantum simulation platform used.

\vspace{1cm}
\emph{Acknowledgements ---} We gratefully acknowledge discussions about this work with M. Sarovar, A. Baczewski, and D. Burgarth. K. R. D., A. H., and C. A. acknowledge support from the National Science Foundation (Grant No. 2231328) and Knowledge Enterprise at Arizona State University. A.B.M. is supported by Sandia National Laboratories’ Laboratory Directed Research and Development Program under the Truman Fellowship. Sandia National Laboratories is a multimission laboratory managed and operated by National Technology \& Engineering Solutions of Sandia, LLC, a wholly owned subsidiary of Honeywell International Inc., for the U.S. Department of Energy’s National Nuclear Security Administration under contract DE-NA0003525. This article has been authored by an employee of National Technology \& Engineering Solutions of Sandia, LLC under Contract No. DE-NA0003525 with the U.S. Department of Energy (DOE). The employee owns all right, title and interest in and to the article and is solely responsible for its contents. The United States Government retains and the publisher, by accepting the article for publication, acknowledges that the United States Government retains a non-exclusive, paid-up, irrevocable, world-wide license to publish or reproduce the published form of this article or allow others to do so, for United States Government purposes. The DOE will provide public access to these results of federally sponsored research in accordance with the DOE Public Access Plan https://www.energy.gov/downloads/doe-public-access-plan. This paper describes objective technical results and analysis. Any subjective views or opinions that might be expressed in the paper do not necessarily represent the views of the U.S. Department of Energy or the United States Government. SAND2024-06421O.

\vspace{1cm}
\noindent

\let\oldaddcontentsline\addcontentsline
\renewcommand{\addcontentsline}[3]{}

\bibliography{Zeno.bib}

\let\addcontentsline\oldaddcontentsline

\onecolumngrid

\newpage 

\begin{center}
\textbf{\large Appendix} 
\end{center}
\appendix

\renewcommand{\appendixname}{}

\tableofcontents

\vspace{1cm}

In this Appendix we begin by reviewing identities and introducing notation. We then derive bounds for the first- and second-order Zeno error and the corresponding success probabilities given in the main body of the manuscript. We also derive bounds for the Zeno error in the case where Zeno-based Hamiltonian simulation is implemented through unitary kicks, and for the diamond distance between the corresponding quantum channels akin QDRIFT.

\section{Identities}

\subsection{Norms}
Throughout this Supplemental Material we consider the spectral norm
\begin{align}
\Vert A\Vert=\sup_{\Vert x\Vert_{2}\neq 0}\frac{\Vert A x\Vert_{2}}{\Vert x\Vert _{2}},
\end{align}
which is induced by the $\ell_2$-norm and is given by the largest eigenvalue of $\sqrt{A^{\dagger}A}$, and the diamond norm 
\begin{align}
\Vert \mathcal E\Vert_{\diamond}=\sup_{\Vert \rho\Vert_{1}=1} \Vert  (\mathcal E\otimes \text{id})(\rho)\Vert_{1},
\end{align}
where $\Vert A \Vert _{1}$ denotes the trace norm. Both norms satisfy the triangle inequality  
$\Vert A+B\Vert \leq \Vert A\Vert +\Vert B\Vert$ and submultiplicativity $\Vert AB\Vert \leq \Vert A\Vert \Vert B\Vert$. 

\subsection{Integral remainder of the Taylor series of an exponential}
Let $X$ be an operator or a superoperator and $\alpha \in \mathbb R$. Then, the remainder of the $k+1$ order of the Taylor series,
\begin{equation}
    e^{\alpha X} = \sum_{n = 0}^{\infty} \frac{\alpha^n}{n!} X^n, \nonumber 
\end{equation}
can be written as 
\begin{equation}\label{eq:a07}
    e^{\alpha X} - \sum_{n = 0}^{k} \frac{\alpha ^n}{n!} X^n = \int_{0}^{\alpha} ds X^{k+1} \frac{(\alpha - s)^{k}}{k!}  e^{sX}.
\end{equation}
This identity can be shown by using partial integration. If we separate the first $k+1$ terms in the Taylor series we get
\begin{equation}
    e^{\alpha X} - \sum_{n = 0}^{k} \frac{\alpha^n}{n!} X^n  = \sum_{n=k+1}^{\infty} \frac{\alpha^n}{n!} X^n = \sum_{n=k+1}^{\infty} \int_{0}^{\alpha} ds \frac{s^{n-1}}{(n-1)!} X^{n} = \sum_{n=k}^{\infty} \int_{0}^{\alpha} ds X \frac{s^{n}}{n!} X^{n}. \nonumber 
\end{equation}
Using the substitutions,
\begin{align*}
    u &= X \frac{s^{n}}{n!} X^{n},  &  du &= X \frac{s^{n-1}}{(n-1)!} X^{n}ds,\\
     v &= s - \alpha,    &    dv &= ds,
\end{align*}
and performing integration by parts we arrive at, 
\begin{equation}
    e^{\alpha X} - \sum_{n = 0}^{k} \frac{\alpha^n}{n!} X^n  = \sum_{n=k-1}^{\infty} \int_{0}^{\alpha} ds X^2  \frac{s^{n}}{n!} X^{n}(\alpha - s). \nonumber 
\end{equation}
Employing integration by parts again using the change of variables,
\begin{align}
    u &= X^2 \frac{s^{n}}{n!} X^{n},  &  du &= X^2 \frac{s^{n-1}}{(n-1)!} X^{n}ds,\nonumber \\
     v &=- \frac{(\alpha - s)^2}{2},   &    dv &= (\alpha - s)ds, \nonumber 
\end{align}
we have 
\begin{equation}
    e^{\alpha X} - \sum_{n = 0}^{k} \frac{\alpha^n}{n!} X^n  = \sum_{n=k-2}^{\infty}  \int_{0}^{\alpha} ds X^3  \frac{s^{n}}{n!} X^{n}\frac{(\alpha - s)^2}{2}. \nonumber
\end{equation}
If we repeat this procedure $k$ times in total, so that the sum on the right-hand side starts at $0$, we obtain the desired result given in Eq. \eqref{eq:a07}.

\subsection{Telescoping identity}
For two operators $A$ and $B$ the telescoping identity \cite{hahn2022unification} holds,
\begin{align}
\label{eq:id1}
A^{N}-B^{N}=\sum_{k=0}^{N-1}A^{k}(A-B)B^{N-1-k},
\end{align}
where $N$ is a non-negative integer. Using submultiplicativity and the triangle inequality, we find, 
\begin{align}
\label{eq:id3}
\Vert A^{N}-B^{N} \Vert\leq \sum_{k=0}^{N-1}\Vert A\Vert^{k}\Vert A-B\Vert \Vert B\Vert^{N-1-k}.
\end{align}

\section{Bounding the first-order Zeno error and success probability}\label{sec:firstorder}

\subsection{Success probability} 
\label{sec:succes}
The success probability $p_{\text{succ}, 0}(\Delta t)=\bra{\varphi(\Delta t)}(\mathds{1}\otimes \ket{0}\bra{0}^{\otimes n_{a}})\ket{\varphi(\Delta t)}$ associated with observing the bit string composed of all $0$'s on the ancilla register, where the state $\ket{\varphi(\Delta t)}$ is given in the main text in Eq. \eqref{eq:stateZeno}, can be written as

\begin{align}
    p_{\text{succ},0}  &= \bra{\psi_{0}}\otimes\bra{\phi} e^{+i\lambda \Tilde{H} \Delta t} \Tilde{P} e^{-i\lambda \Tilde{H} \Delta t} \ket{\psi_{0}} \otimes \ket{\phi}\nonumber \\
    &=\sum_{n=0}^{\infty} \frac{\lambda^n \Delta t^n}{n!} \bra{\psi_{0}}\otimes\bra{\phi} \text{ad}_{\Tilde{H}}^n (\Tilde{P}) \ket{\psi_{0}} \otimes \ket{\phi}, \label{eq:a3}
\end{align}
where $\text{ad}_{\Tilde{H}}(\cdot)=i[\tilde{H},\cdot]$ is the generator of the unitary channel $e^{\lambda\Delta t\,\text{ad}_{\Tilde{H}}}(\tilde{P})=e^{i\lambda \tilde{H}\Delta t} \tilde{P}e^{-i\lambda \tilde{H}\Delta t} $ and we omit the explicit dependence of $p_{\text{succ},0}$ on $\Delta t$. Since the first order in $\Delta t$ vanishes, the success probability becomes, 
\begin{align}
    p_{\text{succ},0} 
    =& 1 + \sum_{n=2}^{\infty} \frac{\lambda^n \Delta t^n}{n!} \bra{\psi_{0}}\otimes\bra{\phi} \text{ad}_{\Tilde{H}}^n (\Tilde{P})  \ket{\psi_{0}} \otimes \ket{\phi},
\end{align}
so that, 
\begin{align}
    p_{\text{succ},0} - 1 &= \sum_{n=2}^{\infty} \frac{\lambda^n \Delta t^n}{n!}  \bra{\psi_{0}}\otimes\bra{\phi} \text{ad}_{\Tilde{H}}^n (\Tilde{P})  \ket{\psi_{0}} \otimes \ket{\phi} = \bra{\psi_{0}}\otimes\bra{\phi} \left[ e^{\lambda \Delta t\,\text{ad}_{\Tilde{H}}} - \mathds{1} -\lambda \Delta t\, \text{ad}_{\Tilde{H}}\right]  (\Tilde{P})  \ket{\psi_{0}} \otimes \ket{\phi}.
\end{align}
Using identity  (\ref{eq:a07}) we obtain, 
\begin{align}
    1-p_{\text{succ},0}&= -\bra{\psi_{0}}\otimes\bra{\phi}\left [ \int_{0}^{\Delta t} ds\,(\Delta t-s)  \lambda^2\,\text{ad}_{\Tilde{H}}^2 e^{\lambda s \,\text{ad}_{\Tilde{H}}}  (\Tilde{P})  \right] \ket{\psi_{0}} \otimes \ket{\phi} \label{eq:a9} \\
    &\leq \norm{\int_{0}^{\Delta t}ds\,(\Delta t-s) \lambda^2 \,\text{ad}_{\Tilde{H}}^2 e^{\lambda s (\text{ad}_{\Tilde{H}})}  (\Tilde{P})}\\
    &\leq \lambda^2\int_{0}^{\Delta t} ds\norm{\text{ad}_{\Tilde{H}}^2 e^{\lambda s \,\text{ad}_{\Tilde{H}}}  (\Tilde{P})} | \Delta t-s |.
\end{align}
 Defining $e^{\lambda s \,\text{ad}_{\Tilde{H}}} (\Tilde{P}) = \Tilde{\rho}(s)$ yields,
\begin{align}
    1-p_{\text{succ},0} &\leq \lambda^2\int_{0}^{\Delta t} ds\norm{\text{ad}_{\Tilde{H}}^2 (\Tilde{\rho}(s))} (\Delta t - s)\\
    &\leq 4\lambda^2\int_{0}^{\Delta t} ds \norm{\Tilde{H}}^{2}  (\Delta t - s) \label{eq:a13} \\
    &\leq 2 \lambda^2 \Delta t^2 ,
\end{align}
where we used the triangle inequality,  submultiplicativity of the spectral norm, and the fact that $\Vert \tilde{\rho}(s)\Vert=1$ to bound, 
\begin{equation}
   \Vert \text{ad}_{\Tilde{H}}^2 (\Tilde{\rho}(s))\Vert = \Vert \Tilde{H}^2 \Tilde{\rho}(s) - 2 \Tilde{H} \Tilde{\rho}(s) \Tilde{H} + \Tilde{\rho}(s) \Tilde{H}^2\Vert\leq 4\Vert \Tilde{H} \Vert^{2},
\end{equation}
noting that since $\Tilde{H} = \sum_j H_j \otimes \ket{j}\bra{j}$, we have,  $\norm{\Tilde{H}} = \text{max}_j {\norm{H_j}}=1$. With $\Delta t=\frac{t}{N}$ we therefore obtain the lower bound,  
\begin{equation}
    p_{\text{succ},0} \geq 1 - 2\frac{\lambda^2t^2}{N^2},
\end{equation}
for the success probability. Since the overall failure probability is equal to $1-p_{\text{succ},0}^{N}\leq N(1-p_{\text{succ},0})$, we have a lower bound for the overall success probability $p_{\text{succ}}=p_{\text{succ,0}}^{N}$, 
\begin{equation}
    p_{\text{succ}} \geq 1 - 2\frac{\lambda^2t^2}{N}.
\end{equation}

\subsection{Gate error}
\label{sec:GateError}

\label{sec:boundingErrorZeno}
The upper bound for the gate error  \eqref{eq:ZenoError} given in the main text directly follows from the Zeno error bound given in \cite{hahn2022unification}, 
\begin{align}
\left \Vert \left(Pe^{-iHt}P\right)^{N}-e^{-iPHPt}P \right\Vert \leq \frac{t^{2}\Vert H\Vert^{2}}{N}.
\end{align}
Since $\Vert \lambda \tilde{H}\Vert\leq \lambda$, the result from the main paper Eq. \eqref{eq:ZenoError} immediately follows. However, as developing an upper bound for the Zeno error for the second order Zeno sequence follows a similar derivation as in \cite{hahn2022unification}, we give here the derivation of the bound for the first-order gate error first. 
The Zeno-based Hamiltonian simulation error is defined by 
\begin{align}
\epsilon=\left\Vert \left(\tilde{P}e^{-i\lambda\tilde{H}\Delta t}\tilde{P}\right)^{N}-\left(e^{-i\lambda \tilde{P}\tilde{H}\tilde{P}\Delta t}\tilde{P}\right)^{N}\right\Vert.
\end{align}
We can identify $A=\tilde{P}e^{-i\lambda\tilde{H}\Delta t}\tilde{P}$ and $B=e^{-i\lambda \tilde{P}\tilde{H}\tilde{P}\Delta t}\tilde{P}$ in the telescoping identity \eqref{eq:id1} to arrive at, 
  \begin{align}
  \label{eq:ep}
  \epsilon \leq N \left\Vert \tilde{P}e^{-i\lambda\tilde{H}\Delta t}\tilde{P}- e^{-i\lambda \tilde{P}\tilde{H}\tilde{P}\Delta t}\tilde{P} \right\Vert,	
  \end{align}
 where we used that $\Vert \tilde{P}e^{-i\lambda\tilde{H}\Delta t}\tilde{P}\Vert\leq 1$ and $\Vert e^{-i\lambda \tilde{P}\tilde{H}\tilde{P}\Delta t}\tilde{P}\Vert\leq 1$, due to submultiplicativity and unitary invariance of the spectral norm. We can rewrite the argument of the norm on the right-hand side of \eqref{eq:ep} and use identity \eqref{eq:a07} to obtain, 
 \begin{align}
   \label{eq:ep2}
 \left\Vert \tilde{P}e^{-i\lambda\tilde{H}\Delta t}\tilde{P}- e^{-i\lambda \tilde{P}\tilde{H}P\Delta t}\tilde{P} \right\Vert&=	\left\Vert \tilde{P}\left(e^{-i\lambda \tilde{H}\Delta t}-\mathds{1}+i\lambda \tilde{H}\Delta t\right) \tilde{P}-\left(e^{-i\lambda \tilde{P}\tilde{H}\tilde{P}\Delta t}-\mathds{1}+i\lambda \tilde{P}\tilde{H}\tilde{P}\Delta t \right)\tilde{P}  \right\Vert \nonumber \\
 &\leq  \lambda^{2}\int_{0}^{\Delta t}\,ds (\Delta t-s)\left(\left\Vert \tilde{P}\tilde{H}^{2}e^{-is \lambda \tilde{H}}\tilde{P} \right\Vert+\left \Vert \tilde{P}\tilde{H}^{2}\tilde{P}e^{-is\lambda \tilde{P}\tilde{H}\tilde{P}} \right\Vert\right). 
 \end{align}
Since
\begin{align}
&\left \Vert \tilde{P}\tilde{H}^{2}\tilde{P}e^{-is\lambda \tilde{P}\tilde{H}\tilde{P}} \right\Vert\leq \Vert \tilde{P}\tilde{H}^{2}\tilde{P} \Vert \leq 1, \\
&
\left\Vert \tilde{P}\tilde{H}^{2}e^{-is \lambda \tilde{H}}\tilde{P} \right\Vert \leq 1,
\end{align}
we arrive at our final result, 
\begin{align}
\epsilon  \leq 2N\lambda^{2}\int_{0}^{\Delta t}\,ds(\Delta t-s)=\frac{t^{2}\lambda^{2}}{N}.	
\end{align}

\section{Bounding the second-order Zeno error and success probability}\label{sec:secondorder}
\subsection{Success probability} 
The success probability of one Zeno step using the second-order Zeno sequence \eqref{eq:secondorder} is given by,
\begin{align}
    p_{\text{succ},0} 
    &= \bra{\psi_{0}}\otimes\bra{\phi} e^{+i\lambda \Tilde{H} \Delta t/2} \Tilde{R} e^{+i\lambda \Tilde{H} \Delta t/2}  \Tilde{P} e^{-i\lambda \Tilde{H} \Delta t /2}
    \Tilde{R}e^{-i\lambda \Tilde{H} \Delta t /2}   \ket{\psi_{0}} \otimes \ket{\phi}\\
    &=\sum_{n,m=0}^{\infty} \frac{\lambda^{n+m} (\Delta t/2)^{n+m}}{n! m!} \bra{\psi_{0}}\otimes\bra{\phi} \text{ad}_{\Tilde{H}}^m\left( \Tilde{R}\, \text{ad}_{\Tilde{H}}^n (\Tilde{P}) \Tilde{R}\right)\ket{\psi_{0}} \otimes \ket{\phi},
\end{align}
The terms that depend on ${\Delta t^0 , \Delta t^1, \Delta t^2}$ corresponds to $n + m = 0 ~(\text{zeroth~order}), n + m = 1 ~(\text{first order}), n + m = 2 ~(\text{second order})$. They are given by:
\begin{align}
    \text{Zeroth order}: & \bra{\psi_{0}} \otimes \bra{\phi} \text{ad}_{\Tilde{H}}^0\bigl( \Tilde{R}\,\text{ad}_{\Tilde{H}}^0(\Tilde{P})\Tilde{R}\bigl) \ket{\psi_{0}} \otimes \ket{\phi} 
    = \bra{\psi_{0}} \otimes \bra{\phi} \Tilde{P} \ket{\psi_{0}} \otimes \ket{\phi} = 1 \nonumber,\\
    \\
    \text{First order}:~&\frac{\lambda \Delta t}{2}\bra{\psi_{0}} \otimes \bra{\phi}\left( \text{ad}_{\Tilde{H}}^1\bigl( \Tilde{R}\,\text{ad}_{\Tilde{H}}^0(\Tilde{P})\Tilde{R})\bigl) + \text{ad}_{\Tilde{H}}^0\bigl( \Tilde{R}(\text{ad}_{\Tilde{H}}^1(\Tilde{P})\Tilde{R})\bigl)\right)  \ket{\psi_{0}} \otimes \ket{\phi} \nonumber \\
    &= \frac{i\lambda \Delta t}{2}\bra{\psi_{0}} \otimes \bra{\phi} \left([\Tilde{H},\Tilde{P}] + \Tilde{R}[\Tilde{H},\Tilde{P}]\Tilde{R} \right)\ket{\psi_{0}} \otimes \ket{\phi} \nonumber\\
    &= \frac{i\lambda \Delta t}{2}\bra{\psi_{0}} \otimes \bra{\phi}\left([\Tilde{H},\Tilde{P}] + [\Tilde{H},\Tilde{P}] \right)\ket{\psi_{0}} \otimes \ket{\phi} = 0 \nonumber,\\
    \\
    \text{Second order}:~& \frac{(\lambda \Delta t /2)^2}{2} \bra{\psi_{0}} \otimes \bra{\phi} \Big(2\,\text{ad}_{\Tilde{H}}^1\bigl( \Tilde{R}(\text{ad}_{\Tilde{H}}^1(\Tilde{P})\Tilde{R})\bigl) + \text{ad}_{\Tilde{H}}^2\bigl( \Tilde{R}(\text{ad}_{\Tilde{H}}^0(\Tilde{P})\Tilde{R})\bigl) \nonumber \\ &+ \text{ad}_{\Tilde{H}}^0\bigl( \Tilde{R}(\text{ad}_{\Tilde{H}}^2(\Tilde{P})\Tilde{R})\bigl) \Big) \ket{\psi_{0}} \otimes \ket{\phi} \nonumber \\
    &= -\frac{(\lambda \Delta t /2)^2}{2} \bra{\psi_{0}} \otimes \bra{\phi} \left(2 \Bigr[ \Tilde{H},\Tilde{R}[\Tilde{H},\Tilde{P}]\Tilde{R}\Bigr] + \Bigr[\Tilde{H},[\Tilde{H}, \Tilde{P}]\Bigr] + \Tilde{R} \Bigr[\Tilde{H},[\Tilde{H},\Tilde{P}]\Bigr] \Tilde{R}\right)\ket{\psi_{0}} \otimes \ket{\phi} \nonumber \\
    &= -(\lambda \Delta t /2)^2 \bra{\psi_{0}} \otimes \bra{\phi}\left( \Bigr[ \Tilde{H},\Tilde{R}[\Tilde{H},\Tilde{P}]\Tilde{R}\Bigr] +  \Bigr[\Tilde{H},[\Tilde{H},\Tilde{P}]\Bigr]\right) \ket{\psi_{0}} \otimes \ket{\phi}  \nonumber\\
    &= -(\lambda \Delta t /2)^2 \bra{\psi_{0}} \otimes \bra{\phi}\left( \Tilde{H}\Tilde{R}[\Tilde{H},\Tilde{P}]\Tilde{R} - \Tilde{R}[\Tilde{H},\Tilde{P}]\Tilde{R} \Tilde{H} + \Bigr[\Tilde{H},[\Tilde{H},\Tilde{P}]\Bigr]\right) \ket{\psi_{0}} \otimes \ket{\phi}\nonumber \\
    &=  -(\lambda \Delta t /2)^2 \bra{\psi_{0}} \otimes \bra{\phi} \left(\Tilde{H}\Tilde{R}[\Tilde{H},\Tilde{P}] - [\Tilde{H},\Tilde{P}]\Tilde{R} \Tilde{H} + \Bigr[\Tilde{H},[\Tilde{H},\Tilde{P}]\Bigr]\right) \ket{\psi_{0}} \otimes \ket{\phi} \nonumber \\
    &=  -(\lambda \Delta t /2)^2 \bra{\psi_{0}} \otimes \bra{\phi} \left(\Tilde{H}(2\Tilde{P}-\mathds{1})[\Tilde{H},\Tilde{P}] - [\Tilde{H},\Tilde{P}](2\Tilde{P}-\mathds{1}) \Tilde{H} + \Bigr[\Tilde{H},[\Tilde{H},\Tilde{P}]\Bigr]\right) \ket{\psi_{0}} \otimes \ket{\phi} \nonumber \\
    &= -2(\lambda \Delta t /2)^2 \bra{\psi_{0}} \otimes \bra{\phi} \left(\Tilde{H}\Tilde{P}[\Tilde{H},\Tilde{P}] - [\Tilde{H},\Tilde{P}] \Tilde{P} \Tilde{H} \right) \ket{\psi_{0}} \otimes \ket{\phi} \nonumber\\
    &= -2(\lambda \Delta t /2)^2 \bra{\psi_{0}} \otimes \bra{\phi} \left(\Tilde{H}\Tilde{P}   \Tilde{H}\Tilde{P} -\Tilde{H}\Tilde{P}\Tilde{H} -\Tilde{H}\Tilde{P}\Tilde{H} + \Tilde{P}\Tilde{H}\Tilde{P}\Tilde{H}\right) \ket{\psi_{0}} \otimes \ket{\phi} = 0\nonumber.
\end{align}
We note that $\Tilde{R}$ or $\Tilde{P}$ acts trivially on $\ket{\phi}$ or $\bra{\phi}$ if it is on the right or left side of each term. Thus, $p_{\text{succ},0} - 1$ is given by,
\begin{align}
     p_{\text{succ},0} - 1=&\sum_{n+m \geq 3} \frac{\lambda^{n+m} (\Delta t/2)^{n+m}}{n! m!} \bra{\psi_{0}}\otimes\bra{\phi} \text{ad}_{\Tilde{H}}^m\left( \Tilde{R} \,\text{ad}_{\Tilde{H}}^n (\Tilde{P}) \Tilde{R}\right)\ket{\psi_{0}} \otimes \ket{\phi} \\
    = &\sum_{m = 0} \frac{\lambda^{m} (\Delta t/2)^{m}}{m!} \bra{\psi_{0}}\otimes\bra{\phi} \text{ad}_{\Tilde{H}}^m\left( \Tilde{R} \sum_{n = 3-m}\frac{\lambda^n(\Delta t/2)^n}{n!}\,\text{ad}_{\Tilde{H}}^n (\Tilde{P}) \Tilde{R}\right)\ket{\psi_{0}} \otimes \ket{\phi} \label{eq:secondsum1}\\ 
    =&\sum_{m = 0} \frac{\lambda^{m} (\Delta t/2)^{m}}{m!} \bra{\psi_{0}}\otimes\bra{\phi} \text{ad}_{\Tilde{H}}^m\left( \Tilde{R} \left( e^{\lambda \Delta t/2 \,\text{ad}_{\Tilde{H}}} - \sum_{n = 0}^{3-m-1}\frac{\lambda^n(\Delta t/2)^n}{n!}\,\text{ad}_{\Tilde{H}}^n \right)(\Tilde{P})\Tilde{R}\right)\ket{\psi_{0}} \otimes \ket{\phi} \label{eq:secondsum2},
\end{align}
where we note that for $m\geq 3$ the second sum in \eqref{eq:secondsum2} is zero, i.e., the limit of the second sum in \eqref{eq:secondsum1} starts at $n=0$.  
Defining 
\begin{align}
    \mathds{J}(\Tilde{P}) &= \sum_{m = 0} \frac{\lambda^{m} (\Delta t/2)^{m}}{m!} \,\text{ad}_{\Tilde{H}}^m\left( \Tilde{R} \left( e^{\lambda \Delta t/2 \,\text{ad}_{\Tilde{H}}} - \sum_{n = 0}^{3-m-1}\frac{\lambda^n(\Delta t/2)^n}{n!}\,\text{ad}_{\Tilde{H}}^n \right)(\Tilde{P})\Tilde{R}\right),
\end{align}
and separating the terms corresponding to $m = 0 , m = 1 , m = 2$ and $m \geq 3$, which we denote by  $\mathds{J}_0, \mathds{J}_1, \mathds{J}_2$ and $\mathcal{J}$, respectively, we obtain,
\begin{align} 
    \mathds{J}(\Tilde{P}) =  \mathds{J}_0(\Tilde{P}) + \mathds{J}_1(\Tilde{P}) + \mathds{J}_2 (\Tilde{P}) + \mathcal{J}(\Tilde{P}),
\end{align}
so that,
\begin{align}
    1-p_{\text{succ},0} &= -\bra{\psi_{0}} \otimes \bra{\phi} \mathds{J} (\Tilde{P})\ket{\psi_{0}} \otimes \ket{\phi}
    \leq \norm{\mathds{J}(\Tilde{P})}
    \leq \norm{\mathds{J}_0(\Tilde{P})} + \norm{\mathds{J}_1(\Tilde{P})} + \norm{\mathds{J}_2(\Tilde{P})} + \norm{\mathcal{J}(\Tilde{P})}\label{eq:b13} . 
\end{align}
In order to bound the spectral norm of each of the terms,  we define,
\begin{equation}
    \mathcal{M}_k (\tilde{P}) = \Tilde{R} \left( e^{\lambda \Delta t/2 \,\text{ad}_{\Tilde{H}}} - \sum_{n = 0}^{k}\frac{\lambda^n(\Delta t/2)^n}{n!}\,\text{ad}_{\Tilde{H}}^n \right)(\Tilde{P})\Tilde{R}.
\end{equation}
Since $\norm{\Tilde{R}} = 1$ we can make use of submultiplicativity of the spectral norm and use the identity (\ref{eq:a07}) to obtain,
\begin{align}
    \norm{\mathcal{M}_k (\tilde{P})} \le \norm{  \int_0^{\Delta t/2} ds \frac{(\lambda \,\text{ad}_{\Tilde{H}})^{k+1}}{k!}(\Delta t/2 -s)^{k}e^{\lambda s \,\text{ad}_{\Tilde{H}}} (\Tilde{P})}.
\end{align}
Defining again $\Tilde{\rho}(s) = e^{\lambda s \,\text{ad}_{\Tilde{H}}}(\Tilde{P})$ and using the triangle inequality we have,
\begin{align} 
    \norm{\mathcal{M}_k (\tilde{P})}  &\le \int_0^{\Delta t/2} ds \frac{\lambda^{k+1}}{k!} |\Delta t/2 -s|^{k} \norm{   \text{ad}_{\Tilde{H}}^{k+1}(\Tilde{\rho}(s))} \label{eq:b117} \\
    &\le \frac{2^{k+1}}{k!} \int_0^{\Delta t/2} ds \lambda^{k+1} (\Delta t/2 -s)^{k} = \frac{(\lambda \Delta t )^{k+1}}{(k+1)!} \nonumber ,
\end{align}
where we used the fact that $\Tilde{\rho}(s)$ has spectral norm one, and that $\text{ad}_{\Tilde{H}}^{k+1}(\tilde{\rho}(s)) $ has $2^{k+1}$ terms, each being products of $\Tilde{H}$ and $\tilde{\rho}(s)$, where the spectral norm of both terms is $1$.

We will continue by calculating the spectral norm for each term individually. For $m=0$ we have, 
\begin{align}
    \norm{\mathds{J}_0(\Tilde{P})} &= \norm{\Tilde{R} \left( e^{\lambda \Delta t/2 \,\text{ad}_{\Tilde{H}}} - \sum_{n = 0}^{2}\frac{\lambda^n(\Delta t/2)^n}{n!}\,\text{ad}_{\Tilde{H}}^n \right)(\Tilde{P})\Tilde{R}} = \norm{\mathcal{M}_2 (\tilde{P})},
\end{align}
which yields with (\ref{eq:b117}) for $k = 2$,
\begin{equation}
    \norm{\mathds{J}_0(\Tilde{P})} \le \frac{(\lambda \Delta t)^3}{6}.
\end{equation}
Similarly, for $\mathds{J}_1(\Tilde{P})$ we find,
\begin{align}
    \norm{\mathds{J}_1(\Tilde{P})} &= \norm{\frac{\lambda \Delta t }{2}\,\text{ad}_{\Tilde{H}} \left( \Tilde{R} \left( e^{\lambda \Delta t/2 \,\text{ad}_{\Tilde{H}}} - \sum_{n = 0}^{1}\frac{\lambda^n(\Delta t/2)^n}{n!}\,\text{ad}_{\Tilde{H}}^n \right)(\Tilde{P})\Tilde{R}\right) }\\
    &= \frac{\lambda \Delta t}{2} \norm{\text{ad}_{\Tilde{H}}(\mathcal{M}_1 (\tilde{P}))},
\end{align}
so that, 
\begin{align}
    \norm{\mathds{J}_1(\Tilde{P})} &= \frac{\lambda \Delta t}{2} \norm{\Tilde{H}\mathcal{M}_1 (\tilde{P})- \mathcal{M}_1 (\tilde{P}) \Tilde{H}} \\
    &\le \lambda \Delta t \norm{\Tilde{H}} \norm{\mathcal{M}_1(\tilde{P})}.
\end{align}
Using inequality (\ref{eq:b117}) for $k = 1$ we arrive at,
\begin{equation}
    \norm{\mathds{J}_1(\Tilde{P})} \le \frac{(\lambda \Delta t)^3}{2}.
\end{equation}
For  $\mathds{J}_2(\Tilde{P})$ we have,
\begin{align}
    \norm{\mathds{J}_2(\Tilde{P})} &= \norm{\frac{(\lambda \Delta t)^2 }{8}\,\text{ad}_{\Tilde{H}}^2 \left( \Tilde{R} \left( e^{\lambda \Delta t/2\, \text{ad}_{\Tilde{H}}} - \sum_{n = 0}^{0}\frac{\lambda^n(\Delta t/2)^n}{n!}\,\text{ad}_{\Tilde{H}}^n \right)(\Tilde{P})\Tilde{R}\right) }\\ &= \frac{(\lambda \Delta t)^2 }{8} \norm{\text{ad}_{\Tilde{H}}^2 (\mathcal{M}_0 (\tilde{P}))}.
\end{align}
Since $\text{ad}_{\Tilde{H}}^2 (\mathcal{M}_0 (\tilde{P}))$ will have 4 terms that are products of $\Tilde{H}$ and $\mathcal{M}_0 (\tilde{P})$, and due to the fact that $\norm{\Tilde{H}} = 1$, we obtain, 
\begin{align}
    \norm{\mathds{J}_2(\Tilde{P})} \le \frac{(\lambda \Delta t)^2}{2} \norm{\mathcal{M}_0(\Tilde{P})} \le \frac{(\lambda \Delta t)^3}{2}.
\end{align}
Finally, for $\norm{\mathcal{J}(\Tilde{P})}$ we have,
\begin{align}
    \norm{\mathcal{J}(\Tilde{P})}&= \left\Vert\sum_{m = 3}^{\infty} \frac{\lambda^{m} (\Delta t/2)^{m}}{m!} \,\text{ad}_{\Tilde{H}}^m\left( \Tilde{R} \left( e^{\lambda \Delta t/2 \,\text{ad}_{\Tilde{H}}}  (\Tilde{P})\right)\Tilde{R}\right)\right\Vert ,
\end{align}
and defining $\Tilde{\rho} = \Tilde{R} \left( e^{\lambda \Delta t/2 \,\text{ad}_{\Tilde{H}}}  (\Tilde{P})\right)\Tilde{R}$ gives,
\begin{align}
    \norm{\mathcal{J}(\Tilde{P})} &= \left\Vert\sum_{m = 3}^{\infty} \frac{\lambda^{m} (\Delta t/2)^{m}}{m!} \,\text{ad}_{\Tilde{H}}^m\left( \Tilde{\rho} \right)\right\Vert .
\end{align}
We now can use identity (\ref{eq:a07}) to obtain, 
\begin{align}
    \norm{\mathcal{J}(\Tilde{P})} &= \norm{\int_{0}^{\Delta t/2} ds (\Delta t/2 - s)^2 \frac{\lambda^3 \,\text{ad}_{\Tilde{H}}^3}{2} e^{\lambda s \,\text{ad}_{\Tilde{H}}}(\Tilde{\rho})}\\
    &\le \frac{\lambda^3}{2}  \int_{0}^{\Delta t/2} ds (\Delta t/2 - s)^2 \norm{\text{ad}_{\Tilde{H}}^3  e^{\lambda s \,\text{ad}_{\Tilde{H}}}(\Tilde{\rho})}.
\end{align}
Since the spectral norm of $\Tilde{\rho}$ is $1$,  we find, 
\begin{align}
    \norm{\mathcal{J}(\Tilde{P})} 
    \le \frac{2^3 \lambda^3}{2}  \int_{0}^{\Delta t/2} ds (\Delta t/2 - s)^2\norm{\Tilde{H}}^{3} \norm{\tilde{\rho}}
    = 4\lambda^3 \int_{0}^{\Delta t/2} ds (\Delta t/2 - s)^2 = \frac{(\lambda \Delta t)^3}{6},
\end{align}
were we again used the fact that $\text{ad}_{\Tilde{H}}^3$  will have $2^3$ terms. In total we arrive at, 
\begin{align}
     p_{\text{succ,0}}  &\geq 1 -  \norm{\mathds{J}_0(\Tilde{P})} - \norm{\mathds{J}_1(\Tilde{P})} - \norm{\mathds{J}_2(\Tilde{P})} - \norm{\mathcal{J}(\Tilde{P})} \\
     &\geq 1 - \frac{4}{3}(\lambda \Delta t)^3.
\end{align}
Analogously to the first-order case, we obtain a lower bound for the total success probability, 
\begin{align}
    p_{\text{succ}} \geq 1 - \frac{4\lambda^3t^3}{3N^2}.
\end{align}
\subsection{Gate error} 

To upper bound the gate error, we first show that the sequence
\begin{equation}\label{eq:b38}
    \Tilde{P}\Tilde{U}(\Delta t/2) \Tilde{R} \Tilde{U}(\Delta t/2) \Tilde{P}, 
\end{equation}
matches
\begin{equation}
    e^{-iH\Delta t}\otimes P,
\end{equation}
up to second order in $\Delta t$. 
From the anti-commutator version of Baker-Campbell-Hausdorff relation \cite{I_Mendas_1989} we can write the sequence as,
\begin{equation}
    \Tilde{P} \sum_{n=0}^{\infty} \frac{\lambda^n (\Delta t/2)^n}{n!} \text{ac}_{\Tilde{H}}^n(\Tilde{R})  \Tilde{P}, 
\end{equation}
where,
\begin{equation}
    \text{ac}_{\Tilde{H}}(\cdot) = -i\{\Tilde{H},\cdot \}. 
\end{equation}
The term corresponding to $\Delta t^0$ is given by,
\begin{equation} 
     \Tilde{P} \Tilde{R}  \Tilde{P} = \mathds{1} \otimes P, \label{0 order} 
\end{equation}
while for $\Delta t^1$ we have,
\begin{equation}
    \frac{-i\lambda \Delta t}{2}  \Tilde{P} \{\Tilde{H},\Tilde{R}\}  \Tilde{P} = -i\lambda\Delta t  \Tilde{P} \Tilde{H} \Tilde{P} = -i \Delta t H \otimes P. \label{1 order}
\end{equation}
For the $\Delta t^2$ term we obtain,
\begin{align}
    \frac{-(\lambda \Delta t)^2}{8}  \Tilde{P} \{\Tilde{H}\{\Tilde{H},\Tilde{R}\}\}  \Tilde{P} &=  \frac{-(\lambda \Delta t)^2}{4}   \Tilde{P} \left( \Tilde{H}^2 + \Tilde{H}\Tilde{R}\Tilde{H}  \right) \Tilde{P}\nonumber \\
    =  \frac{-(\lambda \Delta t)^2}{2} \Tilde{P} \Tilde{H} \Tilde{P} \Tilde{H} \Tilde{P} &= \frac{-\Delta t^2}{2} H^2 \otimes P. \label{2 order}
\end{align}
We thus see that the equations (\ref{0 order}), (\ref{1 order}), and (\ref{2 order}) are the first three terms of the Taylor series of $e^{-iHt}\otimes P$. \\

As before, the gate error
\begin{equation}
    \epsilon =  \Biggr{\|} \left( \Tilde{P}\Tilde{U}(\Delta t/2) \Tilde{R}\;\Tilde{U}(\Delta t/2)  \Tilde{P}\right)^N- \left(e^{-iH \Delta t}\otimes P\right)^{N}\Biggr{\|}, 
\end{equation}
can be upper bounded by using the telescoping identity (\ref{eq:id3}) to arrive at,
\begin{align}
    \epsilon \le   N\Biggr{\|}  \Tilde{P} \sum_{n=0}^{\infty} \frac{\lambda^n (\Delta t/2)^n}{n!} \text{ac}_{\Tilde{H}}^n(\Tilde{R})  \Tilde{P} - e^{-iH \Delta t}\otimes P\Biggr{\|}.
\end{align}
Since we have shown that terms up to second order in $\Delta t$ cancel with the corresponding terms in $e^{-iH \Delta t}\otimes P$, we have,
\begin{align}
    \epsilon \le& N\Biggr{\|}  \Tilde{P} \sum_{n=3}^{\infty} \frac{\lambda^n (\Delta t/2)^n}{n!} \text{ac}^n_{\Tilde{H}}(\Tilde{R})  \Tilde{P} - \left(e^{-iH \Delta t} -\mathds{1} +iH\Delta t +\frac{1}{2}H^2\Delta t^2 \right)\otimes P\Biggr{\|} \nonumber\\
    \le& N \norm{ \sum_{n=3}^{\infty} \frac{\lambda^n (\Delta t/2)^n}{n!} \text{ac}^n_{\Tilde{H}}(\Tilde{R})}   + N \norm{e^{-iH \Delta t} -\mathds{1} +iH\Delta t +\frac{1}{2}H^2\Delta t^2}
    \end{align}

We can further upper bound each term by using the integral remainder of the Taylor series \eqref{eq:a07}. For the first term we have,
\begin{align}
    \norm{ \sum_{n=3}^{\infty} \frac{\lambda^n (\Delta t/2)^n}{n!} \text{ac}^n_{\Tilde{H}}(\Tilde{R})}  &= \norm{\left( e^{\frac{\lambda \Delta t}{2}\text{ac}_{\Tilde{H}} } - \mathds{1} -\frac{\lambda \Delta t}{2} \text{ac}_{\Tilde{H}} - \frac{\lambda^2 \Delta t^2}{8} \text{ac}_{\Tilde{H}}^2 \right)(\Tilde{R})} \nonumber\\
    &\le \int_{0}^{\Delta t/2}ds \lambda^3 \frac{(\Delta t/2 - s)^2}{2} \norm{\text{ac}_{\Tilde{H}}^3 e^{s\lambda \,  \text{ac}_{\Tilde{H}}}(\Tilde{R})} .
\end{align}
Since $\norm{e^{s\lambda \,  \text{ac}_{\Tilde{H}}}(\Tilde{R})}=1$, due to unitary invariance of the spectral norm, noting that  $\text{ac}_{\Tilde{H}}^3$ contains $8$ terms, we find, 
\begin{equation}
    \norm{ \sum_{n=3}^{\infty} \frac{\lambda^n (\Delta t/2)^n}{n!} \text{ac}^n_{\Tilde{H}}(\Tilde{R})} \le 8 \int_{0}^{\Delta t/2}ds \lambda^3 \frac{(\Delta t/2 - s)^2}{2} = \frac{\lambda^3 \Delta t^3}{6}.
\end{equation}
Similarly for the second term, considering that $\norm{H} \le \lambda$ we obtain,
\begin{align}
    \norm{e^{-iH \Delta t} -\mathds{1} +iH\Delta t +\frac{1}{2}H^2\Delta t^2} &\le \int_{0}^{\Delta t}ds \frac{(\Delta t - s)^2}{2} \norm{H^3 e^{-iHs}} \nonumber \\ &\le \int_{0}^{\Delta t}ds \lambda^3 \frac{(\Delta t - s)^2}{2} = \frac{\lambda^3 \Delta t^3}{6}.
\end{align}
For the overall gate error we therefore arrive at our final result,
\begin{equation}
    \epsilon \le N \frac{\lambda^3 \Delta t^3}{3} = \frac{\lambda^3 t^3}{3N^2} \label{2 g error}.
\end{equation}

\section{Zeno-based Hamiltonian simulation through unitary kicks}\label{sec:unitarykicks}
Quantum Zeno dynamics can also be obtained through the fast application of unitary transformations (``kicks''), $U_{Z}$, applied according to the sequence $\left(U_{Z}e^{-iH\Delta t}\right)^{N}\approx U_{Z}^{N} e^{-iH_{Z}t}$ \cite{burgarth2022one}. Here, $H_{Z}=\sum_{l=1}^{m}P_{l}HP_{l}$, where $P_{l}$ are the eigenprojectors of $U_{Z}$ with corresponding (distinct) eigenvalues $\{e^{-i\phi_{l}}\}_{l=1}^{m}$. Thus, if we start in an eigenstate of the kick operator $U_{Z}$, in the limit of infinitely fast kicks the dynamics is governed by the Hamiltonian $H_{Z}$. The corresponding Zeno error in the case of finitely many unitary kicks can be upper bounded by (\cite{burgarth2022one}  Corollary 4),  
\begin{align}
\label{eq:errorboundKicks}
\epsilon=\left \Vert \left(U_{Z}e^{-iH\Delta t}\right)^{N}-U_{Z}^{N}e^{-iH_{Z}t} \right \Vert \leq \frac{2}{N}\left(\frac{\sqrt{m}}{\eta}+1\right)t\Vert H\Vert (1+2t\Vert H\Vert ),
\end{align}
where $\eta=\min_{k\neq l}\left |e^{-i\phi_{k}}-e^{-i\phi_{l}} \right|$.

For Zeno-based Hamiltonian simulation through unitary kicks, we consider the sequence 
$(\tilde{R}\tilde{U}(\Delta t))^{N}$, where $\tilde{U}(\Delta t)=e^{-i\lambda \tilde{H}\Delta t}$, $\tilde{H}=\sum_{j=1}^{L}H_{j}\otimes \ket{j}\bra{j}$, and $\tilde{R}=\mathds{1}\otimes R$. Here, $R=\ket{\phi}\bra{\phi}-\ket{\phi^{\perp}}\bra{\phi^{\perp}}$, where $\ket{\phi^{\perp}}\bra{\phi^{\perp}}$ projects into the subspace orthogonal to the Zeno subspace characterized by $\ket{\phi}\bra{\phi}$. Since $\tilde{R}$ has $m=2$ distinct eigenvalues $1$ and $-1$ with corresponding eigenprojections $\tilde{P}_{1}=\mathds{1}\otimes \ket{\phi}\bra{\phi}$ and $\tilde{P}_{2}=\mathds{1}\otimes \ket{\phi^{\perp}}\bra{\phi^{\perp}}$, respectively, we find

\begin{align}
\lim_{N\to \infty}(\tilde{R}\tilde{U}(\Delta t))^{N}\ket{\psi_{0}}\otimes \ket{\phi}&=\exp(-i\lambda t(\tilde{P}_{1}\tilde{H}\tilde{P}_{1}+\tilde{P}_{2}\tilde{H}\tilde{P}_{2}))\ket{\psi_{0}}\otimes \ket{\phi} \nonumber \\
&=\exp(-it(H\otimes \ket{\phi}\bra{\phi}+H^{\perp}\otimes \ket{\phi^{\perp}}\bra{\phi^{\perp}}))\ket{\psi_{0}}\otimes \ket{\phi}\nonumber \\
&=\exp(-it(H\otimes \ket{\phi}\bra{\phi}))\exp(-it(H^{\perp}\otimes \ket{\phi^{\perp}}\bra{\phi^{\perp}}))\ket{\psi_{0}}\otimes \ket{\phi}
&=e^{-iHt}\ket{\psi_{0}}\otimes \ket{\phi},
\end{align}
where we defined $H^{\perp}=\lambda\sum_{j=1}^{L}c_{j}H_{j}$ with $c_{j}=|\langle j|\phi^{\perp}\rangle|^{2}$. 
Using \eqref{eq:errorboundKicks} and $\Vert \tilde{H}\Vert \leq 1$, the Zeno error can be upper bounded by 
\begin{align}
  \epsilon \leq \frac{2}{N} \left (\frac{1}{\sqrt{2}}+1\right)\lambda t(1+2\lambda t). 
\end{align}

\section{Upper bound for the diamond distance}\label{sec:diamond}
Here we derive an upper bound for the diamond distance, 
\begin{align}
\epsilon_{\diamond}=\left\Vert \mathcal M_{\Delta t}^{N}-\mathcal U_{t}\right\Vert_{\diamond},
\end{align}
between the repeated application of the quantum channel $\mathcal M_{\Delta t}(\cdot)=\sum_{j=1}^{L}p_{j}e^{-\lambda \Delta t \,\text{ad}_{H_{j}}}(\cdot)$ and the unitary channel $\mathcal U_{t}(\cdot)=e^{-t\,\text{ad}_{H}}(\cdot)$. 
Since $\left\Vert \mathcal U_{t}\right\Vert_{\diamond}=1 $ and 
\begin{align}
\left\Vert \sum_{j=1}^{L}p_{j}e^{-\lambda \Delta t\, \text{ad}_{H_{j}}}\right\Vert_{\diamond}\leq \sum_{j=1}^{N}p_{j}\left\Vert e^{-\lambda \Delta t \,\text{ad}_{H_{j}}}\right\Vert_{\diamond}=\sum_{j=1}^{L}p_{j}=1,
\end{align}
from the telescoping \eqref{eq:id1} identity it follows that 
\begin{align}
\epsilon_{\diamond}=N \left \Vert  \mathcal M_{\Delta t}-\mathcal U_{\Delta t} \right\Vert_{\diamond}. 
\end{align}
We recall that both quantum channels $\mathcal M_{\Delta t}$ and $U_{\Delta t}$ match up to first order in  $\Delta t$. If we define 
\begin{align}
\mathcal V_{j}^{(2)}&=\sum_{k=2}^{\infty}\frac{1}{k!}(-\lambda \Delta t)^{k}\,\text{ad}_{H_{j}}^{k},\nonumber\\
\mathcal U^{(2)}&=\sum_{k=2}^{\infty}\frac{1}{k!}(-\lambda \Delta t)^{k}\,\text{ad}_{H}^{k},
\end{align}
we thus have 
\begin{align}
\epsilon_{\diamond}\leq N\left(\sum_{j=1}^{L}p_{j}\Vert \mathcal V_{j}^{(2)}   \Vert_{\diamond} +\Vert \mathcal U^{(2)} \Vert_{\diamond} \right).
\end{align}
Using the integral remainder of the Taylor expansion \eqref{eq:a07} we find 
\begin{align}
\Vert \mathcal V_{j}^{(2)}\Vert_{\diamond}=\lambda^{2}\left  \Vert  \int_{0}^{\Delta t}ds\, (\Delta t-s)\,\text{ad}_{H_{j}}^{2} e^{-s\lambda \,\text{ad}_{H_{j}}}\right \Vert_{\diamond}\leq \frac{\lambda^{2}\Delta t^{2}}{2}\Vert \text{ad}_{H_{j}}^{2}\Vert_{\diamond}\leq 2\lambda^{2}\Delta t^{2} \Vert H_{j}\Vert=2\lambda^{2}\Delta t^{2},
\end{align}
and analogously, $\Vert \mathcal U^{(2)}\Vert_{\diamond}\leq 2\lambda^{2}\Delta t^{2} $. In total, we obtain the final result
\begin{align}
\epsilon_{\diamond}\leq 4\frac{\lambda^{2}t^{2}}{N}. 
\end{align}
\end{document}